\documentclass{llncs}

\usepackage[english]{babel}
\usepackage[utf8x]{inputenc}
\usepackage[T1]{fontenc}

\usepackage{amsmath}
\usepackage{graphicx}
\usepackage[table]{xcolor}
\usepackage[colorlinks=true, allcolors=blue]{hyperref}
\usepackage{multirow}
\usepackage{dingbat}

\definecolor{colorc}{HTML}{c0c0c0}
\definecolor{colora}{HTML}{ff8080}
\definecolor{colorp}{HTML}{47b8b8}
\definecolor{coloro}{HTML}{94bd5e}
\definecolor{colort}{HTML}{99ccff}
\definecolor{colore}{HTML}{eb613d}

\begin{document}

\title{Computational Controversy}
\author{Benjamin Timmermans\inst{1},
Tobias Kuhn\inst{1},
Kaspar Beelen\inst{2},
Lora Aroyo\inst{1}
}

\institute{
Vrije Universiteit Amsterdam
\and
University of Amsterdam
}

\maketitle

\begin{abstract}
Climate change, vaccination, abortion, Trump: Many topics are surrounded by fierce controversies. The nature of such heated debates and their elements have been studied extensively in the social science literature. More recently, various computational approaches to controversy analysis have appeared, using new data sources such as Wikipedia, which help us now better understand these phenomena. However, compared to what social sciences have discovered about such debates, the existing computational approaches mostly focus on just a few of the many important aspects around the concept of controversies. In order to link the two strands, we provide and evaluate here a controversy model that is both, rooted in the findings of the social science literature and at the same time strongly linked to computational methods. We show how this model can lead to computational controversy analytics that cover all of the crucial aspects that make up a controversy.
\end{abstract}

\section{Introduction}

On many topics people from different backgrounds have a shared understanding, or at least have views that are not in contradiction to each other. On some questions, however, like global warming, gun control, the death penalty, abortion, and vaccination, groups of people may strongly disagree despite lengthy interactions and debates~\cite{misra2013topic}. Such situations are commonly called \emph{controversies} and nowadays unfold to a large extent on the Web via different social media, discussion forums, and news platforms. This digital nature has naturally led to many computational approaches to capture and analyze controversy~\cite{awadallah2012opinions,garimella2016quantifying,popescu2010detecting,Borra:2015:SCW:2702123.2702436,dori2015automated,lourentzou2015hotspots}. However, while the social sciences have studied the phenomenon of controversies extensively~\cite{garrison1994changing,martin:controversymanual,clarke:1990controversy,kempner:2011forbidden,horst:2010collective}, there is a lack of a well-founded comprehensive model of controversies for such computational approaches to rely on. For that reason, existing computational approaches have mostly focused on a few hand-picked aspects (such as polarity and emotions), which seems insufficient in the case of the complex and multi-faceted nature of the concept of controversy. To resolve this problem, we present and evaluate here a unified model for controversy, and show how the different relevant aspects can be computationally captured and analyzed.

There are many situations where being able to understand the space of a controversy is essential. For journalists, news agencies and media professionals it is often difficult to present a clear picture of an issue from all perspectives. Governments need to make laws that deal with issues for which it is essential that they have a complete understanding of such issues from an unbiased source. For the general public, understanding a controversy can help prevent a filter bubble, a potentially biased situation where they are only presented with information that they want to see. These problems can be addressed by the computational discovery and analysis of controversies and their elements and aspects.

\section{Controversy, a Disputed Concept?}

\subsection{Explaining Disagreement}

Understanding why societies become divided around specific issues has been a major topic of interest for political scientists, communication specialists and linguists---to name just a few disciplines.
To embed our work within this type of literature, this section reviews some of the crucial concepts that have influenced research on public disputes. 
The following chapter narrows its focus to dissect the ``controversy'' in its constitutive parts.

Communications scientists have explained disagreement in terms of diverging or opposing \emph{frames}. 
According to Gamson and Modigliani \cite{garrison1994changing}: ``[a frame is] a central organizing idea or story line that provides meaning to an unfolding strip of events, weaving a connection among them. The frame suggests what the controversy is, [offering information] about the essence of the issue''. Framing bears on how people perceive issues \textit{and} how they are represented in discourse.
Similar to essentially contested concepts, framing involves selection and salience, i.e. a frame tends to highlight one aspect (or a combination of aspects) at the expense of others. 
Or as Entman argues \cite{entman1993framing}: framing occurs in communication when aspects of a given problem are made more salient, thus promoting a ``particular problem definition, causal interpretation, moral evaluation, and/or treatment recommendations''.
Dardis et al. \cite{dardis2008media} demonstrate how framing affects disagreement by distinguishing between conflict-reinforcing frames, which contain evidence-confirming information, and therefore amplify existing beliefs; and conflict-displacing frames that appeal to both sides of a dispute, and diminish the level of disagreement---changes the adversarial structure of a debate.

Political scientists often invoke the concept of an \emph{ideology} to explain the adversarial positions actors take on public issues. 
Converse \cite{converse1962nature}---in one of the early groundbreaking papers on the topic---describes ideology as a "belief system [...] a configuration of ideas and attitudes in which the elements are bound together by some form of constraint or functional interdependence. 
This line of thinking emphasizes the systemic connections between beliefs. 
For example, it implies that we can predict attitudes toward gun control, given the opinions on abortion and environment.
Freeden \cite{freeden1994political} develops a semantic approach: he perceives concepts, such as ``liberty'' and ``justice'' as the ``building blocks'' of political thought which acquire meaning by virtue of their position within a broader network of ideas: ``ideologies are particular patterned clusters and configurations of political concepts.'' 
The meaning of the concepts, is always relational and contested in nature: ``equality'' and ``social justice'' might be related terms---in the sense that they ``naturally'' imply each other--for a Labour politician, but not for a conservative MP. ``An ideology'', Freeden continues, ``is hence none other than the macroscopic structural arrangement that attributes meaning to a range of mutually defining political concepts''.
Freeden leans heavily on Gallie's \cite{gallie1955essentially} notion of ``Essentially Contested Concepts'' which have the following qualities (1) Appraisive, it signifies a valued achievement (i.e. ``Liberty'') (2) internally complex (3) contains ``rivaling'' descriptions of its component parts (4) Context dependent, can modified in the light of changing circumstances.

Linguists, especially the school of ``Critical Discourse Analysis'' (CDA) pointed to the dialectial relation between language and societal institutions: language use reflects as well as shapes relations of power and dominance, and therefore plays a crucial role in reproducing disagreement.
Ideology, according to this tradition, is defined as common sense, or more precisely as a ``pattern of meaning or frame of interpretation [...] felt to be commonsensical, and often functioning in a normative way'' \cite{verschueren2012ideology}. 
It is composed of ``taken-for-granted'' and therefore unquestioned premisses that are shared within a specific community.
This resembles the Kuhnean scientific paradigm. As \cite{verschueren2012ideology} notices: 
``[paradigms are] specific ways of looking, based on taken-for-granted premises that are shared within a community or generation of scientists.''
Ideological disagreement therefore entails a contestation of these commonsensical norms and prescriptions held by specific segments of society.

Research of structuralist linguists---a movement which was prevalent during the seventies or eighties---attempted to unearth language patterns that elicit or reduce disagreement, by scrutinizing how conflict is initiated (the linguistic or communicational devices used) and how it develops \cite{kakava2001discourse}.  
This boiled down to an analysis of  the structure of arguments and the sequential organization of disagreement. 
Brenneis and Lein, 1977 \cite{brenneis1977you} distinguished three argumentative sequences in role-played disputes among children: repetition, escalation, and inversion. 
In later, cross-cultural, studies they \cite{lein1978children} encountered the same patterns in different countries, but also noted cultural differences related to the tolerance for overlaps and interruptions. 
Boggs \cite{boggs1978development} points to ``contradicting routines'' as the main device for performing disputes. 
Pomerantz \cite{pomerantz1984agreeing} defines ``dispreferred-action'' turn shapes as triggers for dispute. 
These turns contain marked ``dispreference'' features such as ``delays, requests for clarification, partial repeats, and other repair initiators, and turn prefaces''. 
According to Millar et al. \cite{millar1984identifying}, ``three consecutive one-up maneuvers'' serve as a good predictor of verbal conflict: ``a conflict results when speaker B's one-up response to speaker A's one-up statement is responded to with a one-up maneuver by speaker A.''

\subsection{Anatomy of Controversies}

People of different ideologies, seeing the world through different frames, and possibly speaking different languages, thereby become divided by the public debates that are called controversies \cite{martin:controversymanual,clarke:1990controversy,kempner:2011forbidden,horst:2010collective}. The participants in a controversy are typically varied and can be categorized as (1) core-campaigners, (2) occasional campaigners (3) participants encouraged by campaigners (4) sympathisers.
It is through the interaction between core-campaigners and broader sections of the public (termed occasional campaigners and sympathizers) \cite{martin:controversymanual} that such debates spread: Scientific controversies involve non-scientists, as debates are also held outside the scientific laboratories and journals. 
These discussions usually involve (a combination of) several recurring points on which participants disagree, such as benefits, risks, fairness, economics, human rights, decision-making \cite{martin:controversymanual}, but ultimately flow deeper rooted and persistent ideological divisions or opposing value systems~\cite{maynard:1995managing,horst:2010collective}.

Controversies have furthermore the characteristic property that they tend to become unsolvable and persist over time, but nonetheless experience clear punctuations, they ``flare up and die down'', or even follow a cyclical pattern \cite{jasper1988political}. Not only does the intensity of a controversy fluctuate over time, it also follows different rhythms depending on the arena of the debate. Issues can be low-key as a public debate, but heavily disputed among scientists, and of course vice versa. A controversial debate can be held in different platforms (among scientists \cite{kempner:2011forbidden,tarrow:2008polarization} or experts \cite{hallberg1994child}), but usually migrates to the public sphere through the media, through which it engages broad segments of the public \cite{dalgalarrondo2002tragic,jasper1988political}.

Moreover, the increasing delineation of opposing views results in an ever widening disagreement or polarization~\cite{tarrow:2008polarization}: the debate forces  participant to develop coherent viewpoints and manage to navigate a debate by consistently picking the ``right'' side on each of the aspects. Polarization emerges as discussants develop increasingly well-defined but diverging perspectives---a dynamic propelled by core-campaigner who usually develop the templates \cite{martin:controversymanual}.
Given that disputes flow from the beliefs and values participants hold dear, the exchange of opinions is not limited to the ``facts'', but invites strong emotions~\cite{levi:1988psychological}.

\section{Related Work}


In the last few years, many approaches and methods have been proposed to computationally analyze controversies, and many interesting insights have thereby been found.
The OpinioNetIT~\cite{awadallah2012opinions} project, for example, attempts to computationally reconstruct public debates as an exchange of pro and con statements using person-opinion-topic triples. Other work measures the controversy of a topic by building ``conversation graphs'' using a set of Twitter retweets on a given hashtag \cite{garimella2016quantifying}. Another approach uses Twitter to measure the controversy of events \cite{popescu2010detecting}. Their model principally relies on are linguistic, structural and sentiment features. Besides Twitter, Wikipedia has proven useful for modeling controversy on historic data. An example of this is Contropedia, where the metadata associated with Wikipedia pages such as the presence of edits and reverts were used~\cite{Borra:2015:SCW:2702123.2702436}. It has furthermore been shown that controversial pages on the Web can be detected through mapping them to their closest Wikipedia pages \cite{dori2015automated}.

Only a few approaches explicitly tackled the problem of detecting controversy in news articles. \cite{lourentzou2015hotspots} measured which sentences trigger the largest responses in terms of tweets in order to locate the most controversial points in media coverage. \cite{choi2010identifying} identified controversial topics by looking at which ones tend to invoke conflicting sentiment, and \cite{mejova2014controversy} analyzed news using a crowdsourced lexicon that comprises frequent content words for which participants were asked to judge their controversy. Our work aims to put such approaches onto a solid methodological foundation by measuring controversy in a manner that involved all aspects that have been found to be important in the literature on the topic.

\section{Methodology}

Based on the background provided above, we present here our methodology on what we call computational controversy. The main component is our unifying controversy model, which is linked to computational methods to retrieve, capture, and analyze such debates. We also show a generic architecture of how these different aspects can be brought together.

\subsection{The CAPOTE Controversy Model}
Our unifiying model captures the different characteristic aspects of a controversy as identified in the varied literature on the topic.
Based on that, a controversy can be generally defined as a \emph{heated and polarized public debate by a multitude of actors persisting over time}. The key words in this definition that point to the different aspects are ``heated,'' ``polarized,'' ``public,'' ``actors,'' and ``time.'' With some renaming and reordering, this leads us to claim that a Controversy is made from the key aspects of Actors, Polarization, Openness, Time-persistence, and Emotions, which we can show as an informal equation:
\medskip\\
{\small$\mbox{\colorbox{colorc!50}{Controversy}} \sim \mbox{\colorbox{colora!50}{Actors}} + \mbox{\colorbox{colorp!50}{Polarization}} + \mbox{\colorbox{coloro!50}{Openness}} + \mbox{\colorbox{colort!50}{Time-persistence}} + \mbox{\colorbox{colore!50}{Emotions}}$}\medskip\\
As an acronym for this equation, we call our model CAPOTE. The key aspests of controversy are therefore:
\begin{itemize}
\item \textbf{Actors:} A controversy has many participating actors. We wouldn't call it a controversy if it had only a handful of participants.
\item \textbf{Polarization:} Viewpoints are polarized and not uniform or scattered. We call something a controversy only if the participants are grouped in two or more camps that oppose each other, with few people positioning themselves somewhere in between.
\item \textbf{Openness:} A controversy plays out in an open public space, such as the web. We wouldn't call it a controversy if it was all hidden and happening out of sight for society.
\item \textbf{Time-persistence:} A controversy persists over longer stretches of time, typically years or more. A heated debate that is sparked and settled within a single day, for example, would hardly be called a controversy.
\item \textbf{Emotions:} Strong sentiments or emotions are expressed and are an important driver. It is not a controversy if everybody discusses the matter with a cool head and with no personal emotional involvement.
\end{itemize}
Therefore, according to our model and definition, a set of opinions and arguments expressed in a debate can be called a controversy only if all five criteria above are satisfied.
Importantly, all these five aspects can nowadays be algorithmically assessed and quantified based on a variety of techniques and data sources, as we will see below.

\subsection{Computational Controversy}

With modern techniques on natural language processing, machine learning, and network analytics, all five aspects of controversies according to the CAPOTE model can be computationally accessed. The prevalence of the Web furthermore means that most such data are digital-born and relatively easy to retrieve.

The \emph{openness} of a controversy and the generality of the web allows us to use different types of web content mining \cite{liu2004editorial} to retrieve pertinent data in the first place, in the form of newspaper articles, discussions, social media posts, and contents from collaborative platforms like Wikipedia. The openness criterion thereby establishes the entry point for computational controversy analysis.
Based on these data, we can then identify the participating \emph{actors} with techniques including named entity recognition \cite{nadeau2007survey} and social network analysis methods \cite{scott2012social}.
The \emph{emotions} expressed by these actors can furthermore be detected and categorized with a wide array of existing sentiment analysis techniques \cite{pang2008opinion,feldman2013techniques}.
Additionally, we can of course analyse the content of the posts and articles by extracting their topics and involved concepts. For this, we can apply methods such as topic modeling \cite{blei2003latent} and ontology learning \cite{maedche2001ontology}.
These steps may be run independently, or they may depend on each other. For example, the extraction of emotions may depend on the information of extracted actors, or vice versa.

\begin{figure}[tb]
\centering
\includegraphics[width=\textwidth]{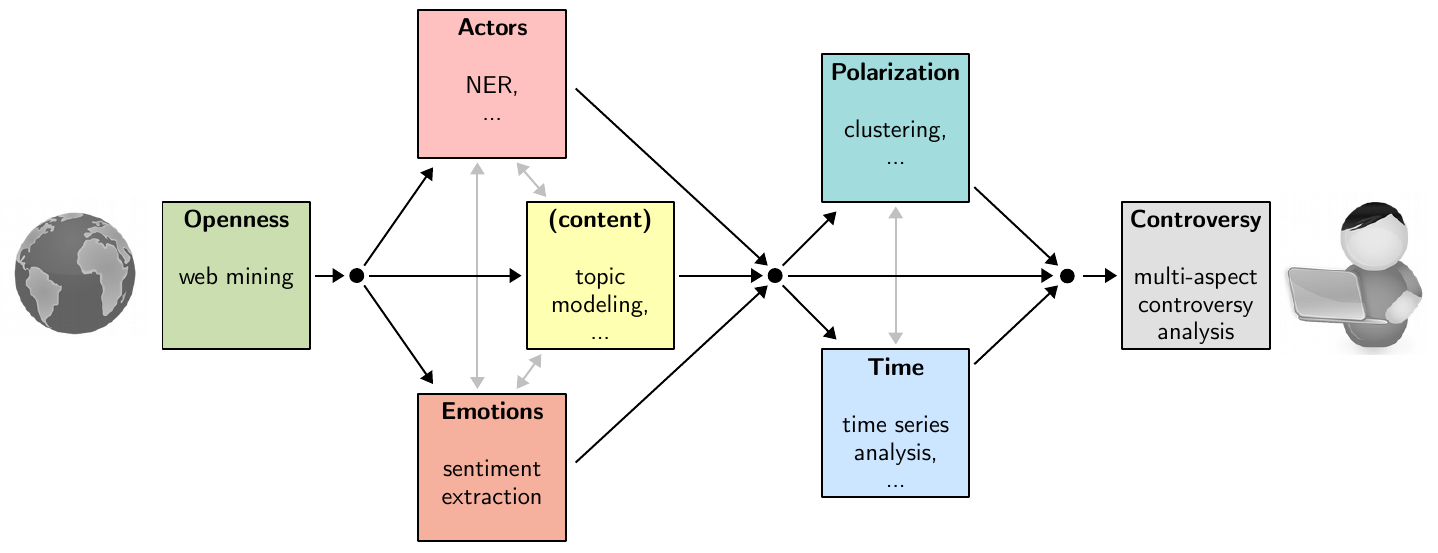}
\caption{A generic CAPOTE-based architecture. The black arrows denote the mandatory data flows for a fully CAPOTE-compliant architecture, whereas the gray ones denote optional data flows.}
\label{fig:capotearch}
\end{figure}

Based on this first round of analysis, we can investigate the remaining aspects of the CAPOTE model.
The \emph{polarization} of viewpoints can be assessed and quantified with clustering and network analysis techniques \cite{andris2015rise}, taking as input the network of actors, their expressions, and the contained topics and emotions.
The \emph{time-persistence} aspect, finally, can be evaluated with time series analyses \cite{box2015time} and dynamic network models \cite{casteigts2012time} on the same input (possibly including polarization and/or feeding its result to the polarization analysis).

All aspects of the CAPOTE model can therefore be extracted with established techniques, and this allows us in the end to combine the results and to analyze controversies in a complete and thorough manner. Figure \ref{fig:capotearch} illustrates this general CAPOTE-based architecture of a system that allows for holistic, multi-aspect controversy analytics.

\section{Evaluation}

We evaluated our approach with a small qualitative study on related work, and a larger quantitative study on the accuracy of our proposed model.

\subsection{Qualitative Study on Related Work}

\begin{table}[tb]
\begin{center}
\caption{Classification of related work with respect to the CAPOTE model.}
\label{tab:relatedworkstudy}
\begin{tabular}{l|>{\columncolor{colora!50}\centering}p{14mm}>{\columncolor{colorp!50}\centering}p{14mm}>{\columncolor{coloro!50}\centering}p{14mm}>{\columncolor{colort!50}\centering}p{14mm}>{\columncolor{colore!50}\centering}p{14mm}l}
Work & Actors & Polarity & Openness & Time & Emotion & \\
\cline{1-6}
Choi et al. (2010) \cite{choi2010identifying} & \checkmark & \checkmark & - & \checkmark & \checkmark & \\
Popescu \& P. (2010) \cite{popescu2010detecting} & \checkmark & - & \checkmark & \checkmark & \checkmark & \\
Awadallah et al. (2012) \cite{awadallah2012opinions} & \checkmark & \checkmark & \checkmark & - & - & \\
Mejova et al. (2014) \cite{mejova2014controversy} & \checkmark & - & \checkmark & - & \checkmark & \\
Borra et al. (2015) \cite{Borra:2015:SCW:2702123.2702436} & \checkmark & \checkmark & - & \checkmark & \checkmark & \\
Dori \& Allan (2015) \cite{dori2015automated} & \checkmark & \checkmark & - & - & - & \\
Lourentzou et al. (2015) \cite{lourentzou2015hotspots} & \checkmark & -  & - & \checkmark & \checkmark & \\
Garimella et al. (2016) \cite{garimella2016quantifying} & \checkmark & \checkmark & \checkmark & - & \checkmark & \\
\end{tabular}
\end{center}
\end{table}

First we start with a small qualitative study of the approaches we introduced as related work on computational controversy analyses. We manually assessed which of the CAPOTE aspects were explicitly considered for each of these works. Table \ref{tab:relatedworkstudy} shows the result.
We see that all existing works on computational controversy cover at least three of our identified CAPOTE aspects, but none covers all five. While the Actors aspect was covered by all, Polarity and Time was covered by most, and Openness and Time was covered only by half of them. In aggregation, these studies had a good coverage, but in isolation each of them missed --- or did not explicitly address --- at least one of the aspects that our literature study identified as a crucial aspect of controversy.

\subsection{Design of Crowd Study}
\label{crowdstudy}

\begin{figure}[tb]
\centering
\includegraphics[width=0.85\textwidth]{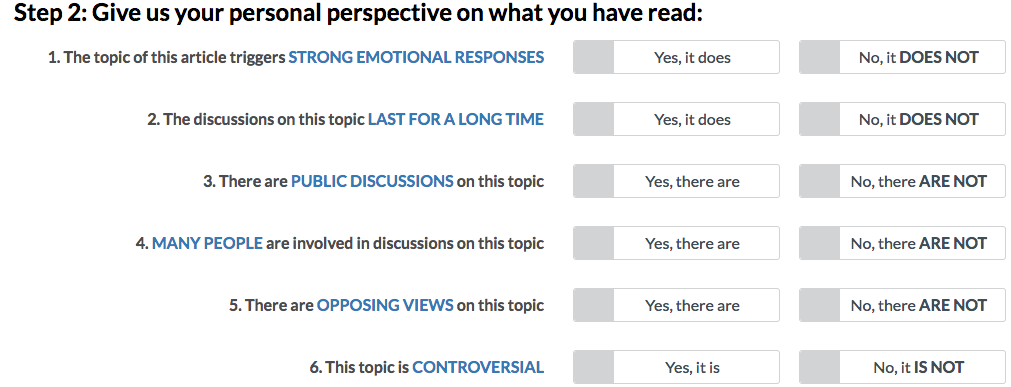}%
\caption{The response section of the user interface of the crowd study}
\label{fig:exp}
\end{figure}

To evaluate the accuracy and completeness of our model we ran as our main study a crowdsourcing experiment using the CrowdFlower\footnote{\url{http://crowdflower.com}} platform. We wanted to find out whether our CAPOTE model aligns with what people would normally call a controversy and thereby whether it is a faithful model of the concept.

To assess the relevance of each of the five aspects, we showed newspaper articles to crowd workers and asked them whether these aspects apply to the given topic and whether they think it deals with a controversy. For this, we showed them the first two paragraphs from $5\,048$ Guardian newspaper articles together with five comments. We retrieved that data through the Guardian news API. Figure \ref{fig:exp} shows the interface with the questions that was shown to the crowd workers. The questions correspond to the five CAPOTE aspects, with an additional question of whether the presented topic was controversial.

The collected annotations from this experiment were evaluated using the CrowdTruth methodology~\cite{aroyo2014three} for measuring the quality of the annotations, the annotators, and the annotated articles. This approach allows the measurement of ambiguity using a vector space. The ambiguity is computed by measuring the cosine distance between vectors of the annotators, where the features or dimensions of the vector represent the possible answers of the annotation task. The same measurement is then used to compare the vector of one annotator to the aggregated vector of all annotators for a single annotated article.


With the resulting data, we are then able to calculate a score between 0 and 1 for each article on these six dimensions, as an average of the workers' ratings. This in turn allows us to run a linear regression analysis to find out about the kind and extent to which the five aspects contribute to the degree to which a given topic is perceived as controversial or not. The five CAPOTE aspects serve as the independent variables in this regression analysis, with the score for conversy serving as the dependent variable to be predicted.



\subsection{Results from Crowd Study}

\begin{table}[tb]
\begin{center}
\caption{Results of the crowdsourcing experiment. For each answer the correlation with the other answers is shown, followed by the ratio of positive answers, the majority vote for yes and the average CrowdTruth relation clarity score as described in section~\ref{crowdstudy}. Below that, the Pearson correlation scores are shown, which was computed by taking the ratio of occurrences for the relation in each article.}
\label{table:results}
\begin{tabular}{ll|cccccc}
 & & \cellcolor{colorc!50} Controversy & \cellcolor{colora!50} Actors & \cellcolor{colorp!50} Polarity & \cellcolor{coloro!50} Openness & \cellcolor{colort!50} Time   & \cellcolor{colore!50} Emotion \\
\hline
\multicolumn{2}{l|}{Ratio of \emph{yes}} & \cellcolor{colorc!50} 0.43 & \cellcolor{colora!50} 0.62 & \cellcolor{colorp!50} 0.57 & \cellcolor{coloro!50} 0.71 & \cellcolor{colort!50} 0.44 & \cellcolor{colore!50} 0.49 \\
\multicolumn{2}{l|}{Majority vote \emph{yes}} & \cellcolor{colorc!50} 0.50 & \cellcolor{colora!50} 0.81 & \cellcolor{colorp!50} 0.73 & \cellcolor{coloro!50} 0.88 & \cellcolor{colort!50} 0.52 & \cellcolor{colore!50} 0.62 \\
\multicolumn{2}{l|}{relation clarity score ~ ~} & \cellcolor{colorc!50} 0.907 & \cellcolor{colora!50} 0.887 & \cellcolor{colorp!50} 0.886 & \cellcolor{coloro!50} 0.915 & \cellcolor{colort!50} 0.883 & \cellcolor{colore!50} 0.890 \\
\hline
correlations: ~ ~ ~ ~ & \cellcolor{colorc!50} C & 1           & 0.4524 & 0.5520   & 0.4655   & 0.5796 & 0.6906 \\
 & \cellcolor{colora!50} A      & 0.4524      & 1      & 0.3848   & 0.5848   & 0.5428 & 0.4618 \\
 & \cellcolor{colorp!50} P    & 0.5520      & 0.3848 & 1        & 0.4067   & 0.3868 & 0.4276 \\
 & \cellcolor{coloro!50} O    & 0.4655      & 0.5848 & 0.4067   & 1        & 0.4564 & 0.4448 \\
 & \cellcolor{colort!50} T        & 0.5796      & 0.5428 & 0.3868   & 0.4564   & 1      & 0.5913 \\
 & \cellcolor{colore!50} E     & 0.6906      & 0.4618 & 0.4267   & 0.4448   & 0.5913 & 1      \\  
\end{tabular}  
\end{center}
\end{table}

In the main experiment first a test was performed on 100 articles to measure how many annotators were required. Each article was annotated by 10 people, after which we found that using six workers would give the best results without significant changes. Following this, a total of $5\,048$ articles were annotated by $1\,659$ unique annotators resulting in a total of $31\,888$ annotations. This dataset is available for download at the CrowdTruth data repository\footnote{\url{http://data.crowdtruth.org}}. Before we turn to the results of the  main linear regression analysis, we can have a look at some descriptive results including Pearson correlation coefficients between the different aspects of the articles.

Table~\ref{table:results} show the results of the descriptive analysis. 43\% of the individual judgment on the overall controversy aspect were positive, leading to a positive controversy classification in 50\% of the articles if a simple majority vote is applied. Out of the five CAPOTE aspects, openness was the most prevalent (71\% of the individual judgments), while time persistence was the least prevalent (44\%). The openness scored highest with .915 for the relation clarity score, which indicates that it is the least ambiguous relation. In contrast, the actors, polarity, time and emotion aspect had similar lower clarity scores, indicating there is more disagreement between the annotators for these relations. The correlation values show the emotion aspect is most strongly correlated with controversy followed by polarity and time persistence, and with actors and openness showing the weakest correlation.

To find out whether these correlations together amplify, we can have a look at the regression results, which are shown in Table \ref{tab:regression}. The top part of the table shows the regression involving all five CAPOTE aspects to predict the controversy aspect. Overall the regression provides a good fit given the inherently noisy nature of human annotations and social science concepts, with an adjusted $R^2$ of 59\%. The effect of all variables is positive, and significant for all of them except Actors. Therefore, we do not have evidence so far that the Actors aspect contributes to the definition of a controversy.

If we look at all combinations of four aspects out of the five, however, we get a more nuanced picture, as shown in the bottom part of Table \ref{tab:regression}. No matter which four aspects we pick, they turn out to be all significant in predicting the controversy of a topic. Therefore, while the Actors aspect does not significantly add to the controversy concept when all other four aspects are present, it does deliver useful redundancy in the sense that it significantly contributes when one of the other aspects is lacking.
These regression analyses furthermore confirm Emotions being the most important aspect. It increases the adjusted $R^2$ by more than 10\%, followed by Polarity, which contributes 5\%, while all other aspects contributing on their own less than 2\%.

\begin{table}[tb]
\begin{center}
\caption{Linear regression analysis on all five aspects (above) and on four of the five aspects (below)}
\label{tab:regression}
\begin{tabular}{l|r>{\columncolor{colora!50}}r>{\columncolor{colorp!50}}r>{\columncolor{coloro!50}}r>{\columncolor{colort!50}}r>{\columncolor{colore!50}}r}
\textbf{\textsc{All 5}} & (intercept) & Actors & Polarity & Openness & Time & Emotions \\
\cline{1-7}
coefficient & ~ -0.15386 & ~~ 0.00787 & ~~ 0.30629 & ~~ 0.10345 & ~~ 0.21832 & ~~ 0.47036 \\
$p$-value & $< 10^{-15}$ & 0.64 & $< 10^{-15}$ & $3.1 \cdot 10^{-10}$ & $< 10^{-15}$ & $< 10^{-15}$ \\
significant & * & & * & * & * & * \\
adjusted $R^2$ & 0.5885 \\
\cline{1-7}
\end{tabular}
\vspace{5mm}\\
\begin{tabular}{l|r>{\columncolor{colora!50}}r>{\columncolor{colorp!50}}r>{\columncolor{coloro!50}}r>{\columncolor{colort!50}}r>{\columncolor{colore!50}}r}
\textbf{\textsc{4 of 5}} & (intercept) & Actors & Polarity & Openness & Time & Emotions \\
\cline{1-7}
coefficient & ~ -0.15267 & & ~~ 0.30687 & ~~ 0.10650 & ~~ 0.22040 & ~~ 0.47095  \\
$p$-value & $< 10^{-15}$ & & $< 10^{-15}$ & $1.7 \cdot 10^{-12}$ & $< 10^{-15}$ & $< 10^{-15}$ \\
significant & * &  & * & * & * & * \\
adjusted $R^2$ & 0.5885 \\
\cline{1-7}
coefficient & ~ -0.09763 & ~~ 0.04465 & & ~~ 0.16910 & ~~ 0.25043 & ~~ 0.53587 \\
$p$-value & $< 10^{-15}$ & 0.012 &  & $< 10^{-15}$ & $< 10^{-15}$ & $< 10^{-15}$ \\
significant & *  & *  &  & *  & *  & *  \\
adjusted $R^2$ & 0.5378 \\
\cline{1-7}
coefficient & ~ -0.12328 & ~~ 0.05008 & ~~ 0.32073 & & ~~ 0.22726 & ~~ 0.48181 \\
$p$-value & $< 10^{-15}$ & 0.00126 & $< 10^{-15}$ & & $< 10^{-15}$ & $< 10^{-15}$ \\
significant & * & * & * &  & * & * \\
adjusted $R^2$ & 0.5848 \\
\cline{1-7}
coefficient & ~ -0.16247 & ~~ 0.07436 & ~~ 0.32261 & ~~ 0.12410 & ~~  & ~~ 0.54804 \\
$p$-value & $< 10^{-15}$ & $6.7 \cdot 10^{-6}$ & $< 10^{-15}$ & $1.3 \cdot 10^{-13}$ &  & $< 10^{-15}$ \\
significant & * & * & * & * &  & * \\
adjusted $R^2$ & 0.5702 \\
\cline{1-7}
coefficient & ~ -0.14464 & ~~ 0.05969 & ~~ 0.39724 & ~~ 0.17575 & ~~ 0.43056 & ~~  \\
$p$-value & $< 10^{-15}$ & 0.00154 & $< 10^{-15}$ & $< 10^{-15}$ & $< 10^{-15}$ &  \\
significant & * & * & * & * & * &  \\
adjusted $R^2$ & 0.4803 \\
\cline{1-7}
\end{tabular}
\end{center}
\end{table}

\section{Conclusions}

Controversies are a frequent and important phenomenon of public discourse. Many approaches have recently been proposed to measure and analyze such controversies with computational means, but a principled framework has been missing. Based on an extensive literature study and supported by a crowdsourced study, we identified five key aspects that define a controversy: a multitude of involved actors, polarized opinions, open visibility of the debate, time persistence, and strong emotions. The results from our crowdsourced study indicate that each of these aspects is a positive indicator of controversy, but also that there is a clear difference in the extend of their influence. Most notably, the emotion aspect was found to be the strongest indicator, while the actors aspect had the weakest influence.

We can often feel that controversies around important issues, such as climate change, are holding us back to make progress on urgent problems. We think that our CAPOTE model can contribute to better understand these controversies and exploit the potential of computational approaches to their analysis. This, in turn, could be the first step towards breaking up the deadlock of long lasting controversial topics.

\section*{Acknowledgments}

This publication was supported by the Dutch national program COMMIT/.

\bibliographystyle{abbrv}
\bibliography{sample}

\end{document}